# Magnetic Space Launcher

*Alexander Bolonkin*
C&R, 1310 Avenue R, #F-6, Brooklyn, NY 11229, USA
(718) 339-4563, aBolonkin@juno.com, http://Bolonkin.narod.ru
*M. Krinker*
Member of Advisory Board of Department of Electrical Engineering,
City College of Technology, CUNY, New York,
mkrinker@aol.com

**Abstract**
A method and facilities for delivering payload and people into outer space are presented. This method uses, in general, engines located on a planetary surface. The installation consists of a space apparatus, power drive stations, which include a flywheel accumulator (for storage) of energy, a variable reducer, a powerful homopolar electric generator and electric rails. The drive stations accelerate the apparatus up to hypersonic speed.

The estimations and computations show the possibility of making this project a reality in a short period of time (for payloads which can tolerate high g-forces). The launch will be very cheap at a projected cost of 3 – 5 dollars per pound. The authors developed a theory of this type of the launcher.

## 1. Introduction

At present, rockets are used to carry people and payloads into space, or to deliver bombs over long distances. This method is very expensive, and requires a well-developed industry, high technology, expensive fuel, and complex devices.

Other than rockets, methods to reach the space velocities are the space elevator, the hypersonic tube air rocket, cable space accelerator, circle launcher and space keeper, centrifugal launcher [1-9], electrostatic liner accelerator [10]. Several new non-rockets methods were also proposed by one of author at the World Space Congress-2002, Houston, USA, 10–19 October 2002.

The space elevator requires very strong nanotubes, as well as rockets and high technology for the initial development. The tube air rocket and non-rocket systems require more detailed research. The electromagnetic transport system, suggested by Minovich (US Patent, 4,795,113, 3 January, 1989)[11], is not realistic at the present time. It requires a vacuum underground tunnel 1530 kilometers long located at a depth of 40 kilometers. The project requires a power cooling system (because the temperature is very high at this depth), a complex power electromagnetic system, and a huge impulse of energy that is greater than the energy of all the electric generating stations on Earth.

This article suggests a very simple and inexpensive method and installation for launching into space.

This is a new space launcher system for delivering hypersonic speeds. This method uses a homopolar electric generator, any conventional power engines (mechanical, electrical, gas turbines), and flywheels (as storage energy) conveniently located on the ground where suspension of weight is not a factor.
**General information about previous works regarding to our topic**.
Below is common information useful for understanding proposed ideas and research.
A **rocket** is a vehicle, missile or aircraft which obtains thrust by the reaction to the ejection of fast moving fluid from within a rocket engine. Chemical rockets operate due to hot exhaust gas made from "propellant" acting against the inside of an expansion nozzle. This generates forces that both accelerate the gas to extremely high speed, as well as, since every action has an equal and opposite reaction, generating a large thrust on the rocket.

The history of rockets goes back to at least the 13th century, possibly earlier. By the 20th century it included human spaceflight to the Moon, and in the 21st century rockets have enabled commercial space tourism.
Rockets are used for fireworks and weaponry, as launch vehicles for artificial satellites, human spaceflight and exploration of other planets. While they are inefficient for low speed use, they are, compared to other propulsion systems, very lightweight, enormously powerful and can achieve extremely high speeds.
Chemical rockets contain a large amount of energy in an easily liberated form, and can be very dangerous, although careful design, testing, construction and use can minimise the risks.
A rocket engine is a jet engine that takes all its reaction mass ("*propellant*") from within tankage and forms it into a high speed jet, thereby obtaining thrust in accordance with Newton's third law. Rocket engines can be used for spacecraft propulsion as well as terrestrial uses, such as missiles. Most rocket engines are internal combustion engines, although non combusting forms also exist.

**Railgun.** Scientists use a railgun for high acceleration of a small conducting body. A railgun is a form of gun that converts electrical energy (rather than the more conventional chemical energy from an explosive propellant) into projectile kinetic energy. It is not to be confused with a coilgun (Gauss gun). Rail guns use magnetic force to drive a projectile. Unlike gas pressure guns, rail guns are not limited by the speed of sound in a compressed gas, so they are capable of accelerating projectiles to extremely high speeds (many kilometers per second).
A wire carrying an electrical current, when in a magnetic field, experiences a force perpendicular to the direction of the current and the direction of the magnetic field.
In an electric motor, fixed magnets create a magnetic field, and a coil of wire is carried upon a shaft that is free to rotate. An electrical current flows through the coil causing it to experience a force due to the magnetic field. The wires of the coil are arranged such that all the forces on the wires make the shaft rotate, and so the motor runs.

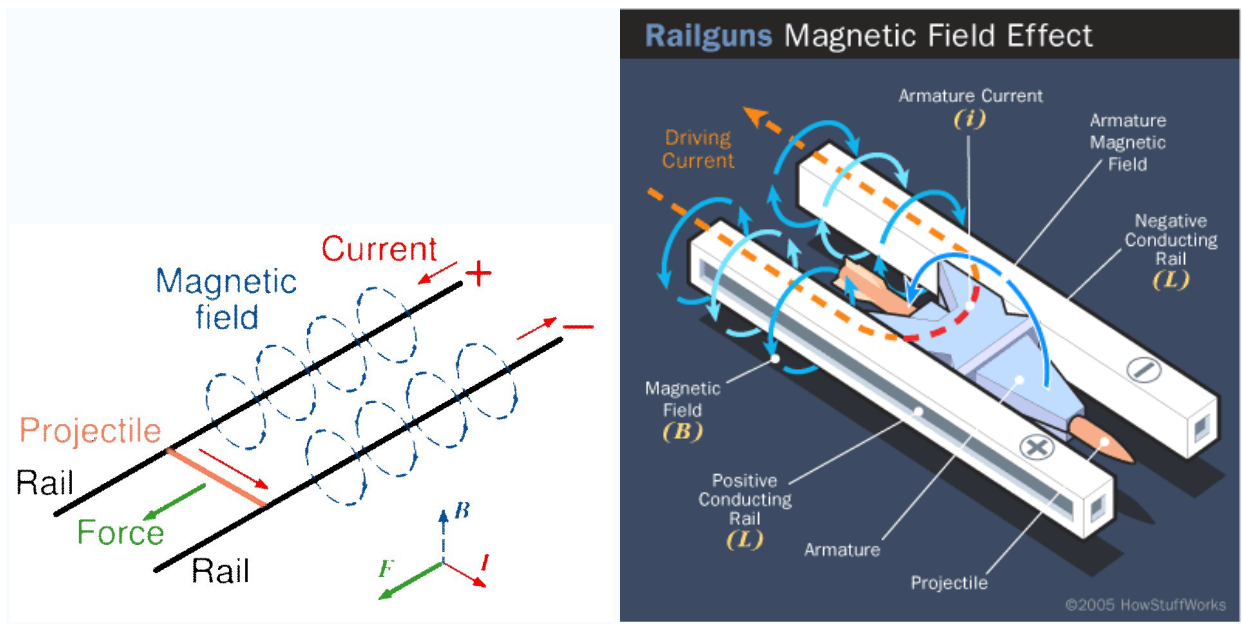

**Fig.1**. Schematic diagrams of a railgun.

A railgun consists of two parallel metal rails (hence the name) connected to an electrical power supply. When a conductive projectile is inserted between the rails (from the end connected to the power supply), it completes the circuit. Electrical current runs from the positive terminal of the power supply up the positive rail, across the projectile, and down the negative rail, back to the power supply (Fig.1).

This flow of current makes the railgun act like an electromagnet, creating a powerful magnetic field in the region of the rails up to the position of the projectile. In accordance with the right-hand rule, the created magnetic field circulates around each conductor. Since the current flows in opposite direction along each rail, the net magnetic field between the rails (**B**) is directed vertically. In combination with the current (**I**) flowing across the projectile, this produces a Lorentz force which accelerates the projectile along the rails. The projectile slides up the rails away from the end with the power supply.

If a very large power supply providing a million amperes or so of current is used, then the force on the projectile will be tremendous, and by the time it leaves the ends of the rails it can be travelling at many kilometres per second. 20 kilometers per second has been achieved with small projectiles explosively injected into the railgun. Although these speeds are theoretically possible, the heat generated from the propulsion of the object is enough to rapidly erode the rails. Such a railgun would require frequent replacement of the rails, or use a heat resistant material that would be conductive enough to produce the same effect.

The need for strong conductive materials with which to build the rails and projectiles; the rails need to survive the violence of an accelerating projectile, and heating due to the large currents and friction involved act against the longevity of the system. The force exerted on the rails consists of a recoil force - equal and opposite to the force propelling the projectile, but along the length of the rails (which is their strongest axis) - and a sideways force caused by the rails being pushed by the magnetic field, just as the projectile is. The rails need to survive this without bending, and thus must be very securely mounted.

The power supply must be able to deliver large currents, with both capacitors and compulsators being common.

The rails need to withstand enormous repulsive forces during firing, and these forces will tend to push them apart and away from the projectile. As rail/projectile clearances increase, electrical arcing develops, which causes rapid vaporization and extensive damage to the rail surfaces and the insulator surfaces. This limits most research railguns to one shot per service interval.

Some have speculated that there are fundamental limits to the exit velocity due to the inductance of the system, and particularly of the rails; but United States government has made significant progress in railgun design and has recently floated designs of a railgun that would be used on a naval vessel. The designs for the naval vessels, however, are limited by their required power usages for the magnets in the rail guns. This level of power is currently unattainable on a ship and reduces the usefulness of the concept for military purposes.

Massive amounts of heat are created by the electricity flowing through the rails, as well as the friction of the projectile leaving the device. This leads to three main problems: melting of equipment, safety of personnel, and detection by enemy forces. As briefly discussed above, the stresses involved in firing this sort of device require an extremely heat-resistant material. Otherwise the rails, barrel, and all equipment attached would melt or be irreparably damaged. Current railguns are not sufficiently powerful to create enough heat to damage anything; however the military is pushing for more and more powerful prototypes. The immense heat released in firing a railgun could potentially injure or even kill bystanders. The heat released would not only be dangerous, but easily detectable. While not visible to the naked eye, the heat signature would be unmistakable to infrared detectors. All of these problems can be solved by the invention of an effective cooling method.

Railguns are being pursued as weapons with projectiles that do not contain explosives, but are given extremely high velocities: 3500 m/s (11,500 ft/s) or more (for comparison, the M16 rifle has a muzzle speed of 930 m/s, or 3,000 ft/s), which would make their kinetic energy equal or superior to the energy yield of an explosive-filled shell of greater mass. This would allow more ammunition to be carried and eliminate the hazards of carrying explosives in a tank or naval weapons platform. Also, by firing at higher velocities railguns have greater range, less bullet drop and less wind drift, bypassing the inherent cost and physical limitations of conventional firearms - "*the limits of gas expansion prohibit launching an unassisted projectile to velocities* *greater than about 1.5 km/s and ranges of more than 50 miles [80 km] from a practical conventional gun system."*

If it is even possible to apply the technology as a rapid-fire automatic weapon, a railgun would have further advantages in increased rate of fire. The feed mechanisms of a conventional firearm must move to accommodate the propellant charge as well as the ammunition round, while a railgun would only need to accommodate the projectile. Furthermore, a railgun would not have to extract a spent cartridge case from the breech, meaning that a fresh round could be cycled almost immediately after the previous round has been shot.

**Tests of Railgun.** Full-scale models have been built and fired, including a very successful 90 mm bore, 9 MJ (6.6 million foot-pounds) kinetic energy gun developed by DARPA, but they all suffer from extreme rail damage and need to be serviced after every shot. Rail and insulator ablation issues still need to be addressed before railguns can start to replace conventional weapons. Probably the most successful system was built by the UK's Defence Research Agency at Dundrennan Range in Kirkcudbright, Scotland. This system has now been operational for over 10 years at an associated flight range for internal, intermediate, external and terminal ballistics, and is the holder of several mass and velocity records.

The United States military is funding railgun experiments. At the University of Texas at Austin Institute for Advanced Technology, military railguns capable of delivering tungsten armor piercing bullets with kinetic energies of nine million joules have been developed. Nine mega-joules is enough energy to deliver 2 kg of projectile at 3 km/s - at that velocity a tungsten or other dense metal rod could penetrate a tank.

The United States Naval Surface Warfare Center Dahlgren Division demonstrated an 8 mega-joule rail gun firing 3.2 kilogram (slightly more than 7 pounds) projectiles in October of 2006 as a prototype of a 64 mega-joule weapon to be deployed aboard Navy warships. Such weapons are expected to be powerful enough to do a little more damage than a BGM-109 Tomahawk missile at a fraction of the projectile cost.

Due to the very high muzzle velocity that can be attained with railguns, there is interest in using them to shoot down high-speed missiles.

**A homopolar generator** is a DC electrical generator that is made when a magnetic electrically conductive rotating disk has a different magnetic field passing through it (it can be thought of as slicing through the magnetic field). This creates a voltage and current difference between 2 contact points, one in the center of the disk the other on the outside of the disk. For simplicity one contact point can be considered positive +, and the other contact point can be considered ground or negative (- or 0). In general the 2 contact points are linked together as the armature. It has the same polarity at every point, so that the armature that passes through the magnetic field lines of force continually move in the same direction. The device is electrically symmetrical (bidirectional), and generates continuous direct current. It is also known as a unipolar generator, acyclic generator, disk dynamo, or Faraday disk (Fig.2). Relatively speaking they can source tremendous electric current (10 to 10000 amperes) but at low potential differences (typically 0.5 to 3 volts). This property is due to the fact that the homopolar generator has very low internal resistance.

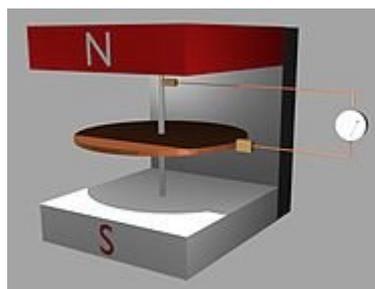

**Fig.2.** Basic Faraday disc generator

The device consists of a conducting flywheel rotating in a magnetic field with one electrical contact near the axis and the other near the periphery. It has been used for generating very high currents at low voltages in applications such as welding, electrolysis and railgun research. In pulsed energy applications, the angular momentum of the rotor is used to store energy over a long period and then release it in a short time.

One of the larger homopolar generators that was produced by Parker Kinetic Designs via the collaboration of Richard Marshall, William Weldon, and Herb Woodson. Parker Kinetic Designs have produced devices which can produce five megaamperes. Another large homopolar generator was built by Sir Mark Oliphant at the Research School of Physical Sciences and Engineering, Australian National University. It produced 500 megajoules and was used as an extremely high-current source for experimentation from 1962 until it was disassembled in 1986. Oliphant's construction was capable of supplying currents of up to 2 megaamperes.

**Magnets**.

**Neodymium magnets** are very strong relative to their mass, but are also mechanically fragile. High-temperature grades will operate at up to 200 and even 230°C but their strength is only marginally greater than that of a samarium-cobalt magnet. As of 2008 neodymium magnets cost about $44/kg, $1.40 per BHmax.

Most neodymium magnets are anisotropic, and hence can only be magnetised along one direction although B10N material is isotropic. During manufacture fields of 30-40 kOe are required to saturate the material. Neodymium magnets have a coercivity (required demagnetisation field from saturation) of about 10,000-12,000 Oersted. Neodymium magnets (or "neo" as they are known in the industry) are graded in strength from N24 to the strongest, N55. The theoretical limit for neodymium magnets is grade N64. The number after the N represents the magnetic energy product, in megagauss-oersteds (MGOe) (1 MG·Oe = 7,958·10³ T·A/m = 7,958 kJ/m³). N48 has a remnant static magnetic field of 1.38 teslas and an *H* (magnetic field intensity) of 13,800 Oersteds (1.098 MA/m). By volume one requires about 18 times as much ceramic magnetic material for the equivalent magnet lifting strength, and about 3 to 5 times as much for the equivalent dipole moment. A neodymium magnet can hold up to 1300 times its own weight.

The neodymium magnet industry is continually working to push the maximum energy product (strength) closer to the theoretical maximum of 64 MGOe. Scientists are also working hard to improve the maximum operating temperature for any given strength.

*Physical and mechanical properties*: Electrical resistivity 160 μ-ohm-cm/cm$^2$; Density 7.4-7.5 g/cm$^3$; Bending strength 24 kg/mm$^2$; Compressive strength 80 kg/mm$^2$; Young's modulus $1.7 \times 10^4$ kg/mm$^2$; Thermal conductivity 7.7 kcal/m-h-°C; Vickers hardness 500 – 600.

**Samarium-cobalt** magnets are primarily composed of samarium and cobalt. They have been available since the early 1970s. This type of rare-earth magnet is very powerful, however they are brittle and prone to cracking and chipping. Samarium-cobalt magnets have Maximum Energy Products (BHmax) that range from 16 Mega-Gauss Oersteds (MGOe) to 32 MGOe, their theoretical limit is 34 MGOe. Samarium Cobalt magnets are available in two "series", namely Series 1:5 and Series 2:17.

*Material properties:* Density: 8.4 g/cm³ ; Electrical Resistivity $0.8 \times 10^{-4}$ Ω·cm; Coefficient of thermal expansion (perpendicular to axis): 12.5 μm/(m·K) .

**Alnico** is an acronym referring to alloys which are composed primarily of aluminium (symbol Al), nickel (symbol Ni) and cobalt (symbol Co), hence *al-ni-co*, with the addition of iron, copper, and sometimes titanium, typically 8–12% Al, 15–26% Ni, 5–24% Co, up to 6% Cu, up to 1% Ti, and the balance is Fe. The primary use of alnico alloys is magnet applications.

Alnico remanence (**B**$_r$) may exceed 12,000 G (1.2 T), its coercion force (**H**$_c$) can be up to 1000 oersted (80 kA/m), its energy product ((**BH**)$_{max}$) can be up to 5.5 MG·Oe (44 T·A/m)—this means alnico can produce high magnetic flux in closed magnetic circuit, but has relatively small resistance against demagnetization.

As of 2008, Alnico magnets cost about $20/pound or $4.30/**BH**$_{max}$.

## 2. Description of Suggested Launcher

**Brief Description**. The installation includes (see notations in Figs. 3, 4): a gun, two electric rails 2, a space apparatus 3, and a drive station 4 (fig.3). The drive station includes: a homopolar electric generator 1 (fig. 4), a variable reducer 3, a fly-wheel energy storage 5, an engine 6, and master drive clutches 2, 4, 6.

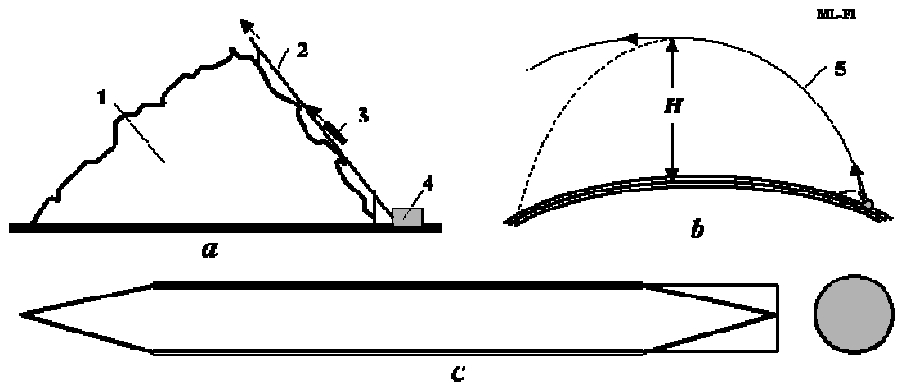

**Fig.3**. Magnetic Launcher. (*a*) Side view; (*b*) Trajectory of space apparatus; (*c*) Hypersonic apparatus. *Notations*: 1 – hill (side view); 2 – railing; 3 – shell; 4 – drive station; 5 – space trajectory.

The system works in the following way:
The engine 7 accelerates the flywheel 5 to maximum safe rotation speed. At launch time, the fly wheel connects through the variable reducer 3 to the homopolar electric generator 1 which produces a high-amperage current. The gas gun takes a shot and accelerates the space apparatus "c" (fig.3) up to the speed of 1500 – 2000 m/s. The apparatus leaves the gun and gains further motion on the rails 2 (fig. 3, fig. 4d) where its body turns on the heavy electric current from the electric generator. The magnetic force of the electric rails accelerates the space apparatus up to speeds of 8000 m/s. (or more) The initial acceleration with a gas gun can decrease the size and cost of the installation when the final speed is not high. The gas gun cannot produce a projectile speed of more than about 2000 m/s. The railgun does not have this limit, but produces some engineering problems such as the required short (pulsed) gigantic surge of electric power, sliding contacts for some millions of amperes current, storage of energy, etc.

The current condensers have a small electric capacity 0.002 MJ/kg ([2], p.465). We would need about $10^{10}$ J energy and 5000 tons of these expensive condensers. The fly-wheels made of cheap artificial fiber have capacity about 0.5 MJ/kg ([2], p.464). The need mass of fly-wheel is decreased to a relatively small 25 – 30 tons. The unit mass of a fly-wheel is significantly cheaper then unit mass of the electric condenser.

The offered design of the magnetic launcher has many innovations which help to overcome the obstacles afforded by a conventional railgun. Itemizing some of them:

1. Fly-wheels from artificial fiber.
2. Small variable reducer with smooth change of turns and high variable rate.
3. Multi-stage monopolar electric generator having capacity of producing millions of amperes and a variable high voltage during a short time.
4. Sliding mercury (gallium) contact having high pass capacity.
5. Double switch having high capacity and short time switching.
6. Special design of projectile (conductor ring) having permanent contact with electric rail.
7. Thin (lead) film on projectile contacts that improve contact of projectile body and the conductor rail.

8. Homopolar generator has magnets inserted into a disk (wheel) form. That significantly simplifies the electric generator.
9. The rails and electric generator can have internal water-cooling.
10. The generator can return rotation energy back to a flywheel after shooting, while rails can return the electromagnetic energy to installation. That way a part of shot energy may be returned. This increases the coefficient of efficiency of the launch installation.

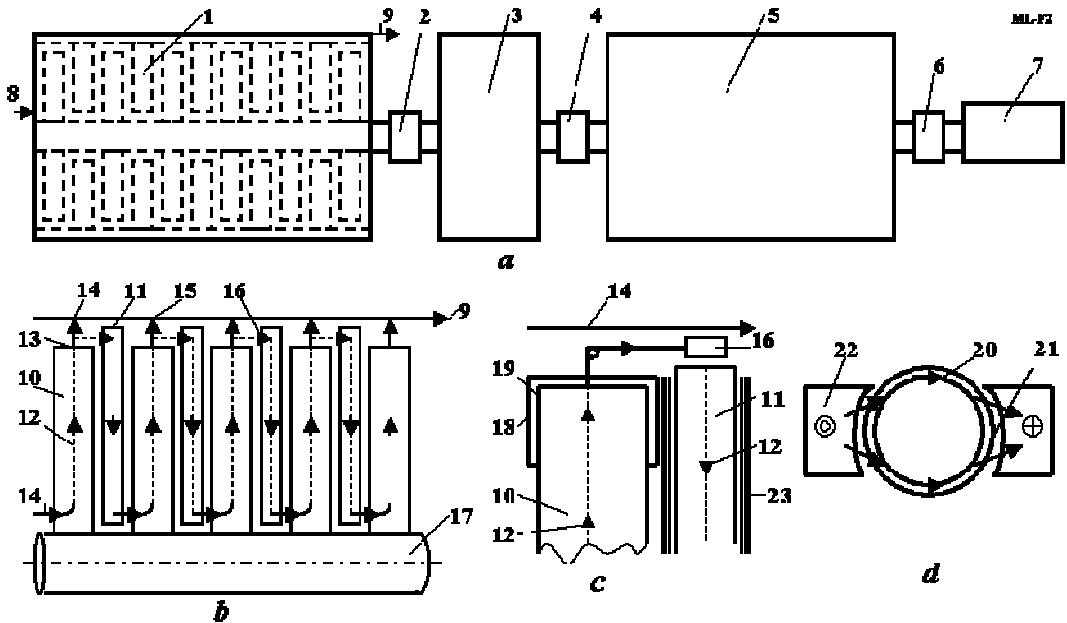

**Fig.3**. Drive station. (*a*) Main components of drive station; (*b*) Rotors and connection disks (wheels); (*c*) Association of rotor and connection disk; (*d*) Association of shell and electric rails (plough or sled). *Notations:* 1 – Electric homogenerator; 2, 4, 6 – master drive clutch; 3 – variable reducer; 5 – fly-wheel; 7 – engine; 8 – enter of electric line; 9 – exit of electric line; 10 – disk (wheel) of rotor (rigid attachment to shaft 17); 11 – motionless conductor (rigid attachment to stator); 12 – electric current; 13 – sliding contact; 14, 15 – exit conductor; 16 – double switch from electric line 14 to conductor 11; 18 – sliding contact; 19 – mercury; 20 – electric ring; 21 – thin film; 22 – electric rail.

The fly-wheel has a disadvantage in that it decreases its' turning speed when one spends its energy. The prospective space apparatus and space launcher needs, on the contrary, an increase of voltage for accelerating the payload. The homopolar generator really would like to increase the number of revolutions thus increasing the voltage. The offered variable reducer approaches this ideal, keeping constant or even increasing the speed of rotation of the electric generator. In addition, the multi-stage electric generator can additionally increase its' voltage by chaining (concatenation of turning on in series mode) its stages or sections.

The sketch of the variable reducer is shown in fig.5. The tape (inertial transfer roll) 3 rotates from shaft 1 (electric generator) to shaft 2 (fly-wheel). In starting position the tape (roll). diameter $d_1$ of shaft 1 is big while the tape (roll) diameter $d_2$ of the fly-wheel is small and rotation speed of electric generator is small. During the rotation, the tape (roll) diameter of shaft 1 decreases, while the corresponding diameter around shaft 2 increases and the rotation speed of the electric generator increases (assuming a correct design of the reducer). The total change of the rotation speed is $(d_1/d_2)^2$. For example, if $d_1/d_2 = 7$, the total change of rotary speed is 49. This way the rotation speed of the electric generator either increases or stays constant in spite of the fact that the rotary speed of the flywheel is decreasing. The multi-stage electric generator achieves the additional increasing of voltage. Its' sections turn on in series.

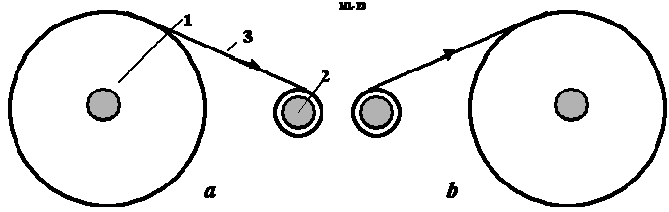

**Fig.5.** Variable reducer. (*a*) Start position; (*b*) final position. *Notations*: 1 – shift of electric generator; 2 – shift of fly-wheel; 3 – tape (inertial transfer roll).

## Theory and computations. Project.

Below is advanced theory of the magnetic launcher and a computation of a sample project.

Let us take the mass of a space apparatus payload $m$ = 150 kg, speed after gas gun $V_o$ = 1500 m/s and final speed $V$ = 8000 m/s.

1. *Estimation of gas gun and magnet accelerator*. Let us take the length of gun barrel $l$ = 15 m. Then average projectile acceleration is

$$a = \frac{V_o^2}{2l} = \frac{(1500)^2}{2 \cdot 15} = 7.5 \cdot 10^4 \quad \text{m/s}^2 . \tag{1}$$

Let us take this acceleration as constant value for main rail magnet acceleration. Then time $t$, length $L$ of magnetic acceleration, force $F$, and energy $A$ are

$$V = V_o + at; \quad t = (V - V_o)/a = 8.67 \cdot 10^{-2} \text{ s}, \quad L = V_o t + at^2/2 = 412 \text{ m},$$
$$F = ma = 1.12 \cdot 10^7 \text{ N}, \quad A = mV^2/2 = 4.8 \cdot 10^9 \text{ J} \tag{2}$$

2. *Requested electric current* and maximal opposed electric intensity from space ship acceleration.

Let us take the distance between rail $d$ = 0.2m and semi-thickness of rail $a_1$ = 0.05m. The request electric currency $i$ and maximum opposed electric intension $E_m$ are

$$i = \left(\frac{\pi F}{\mu_o}(\ln\left|\frac{d-a_1}{a_1}\right|)^{-1}\right)^{0.5} = 5 \cdot 10^6 \text{ A}, \quad E_m = \frac{\mu_o}{\pi} iV \ln\left|\frac{d-a_1}{a_1}\right| = 17.6 \cdot 10^3 \text{ V}, \tag{3}$$

where $\mu_o = 4\pi \times 10^{-7}$ – magnet constant.

3. *Resistance of electric rails*. Let us consider a copper rail with safety limit of temperature increase $\Delta T$ = 200K. The copper has electric resistance $\rho$ = 1.7×10$^{-8}$ Ω.m specific density $\gamma$ = 8430 kg/m$^3$, heat capacity $c_p$ = 90 J/kg.K. From the equation of heat balance we have need of the cross-section area $s$ of rail:

$$s = i\sqrt{\frac{\rho t}{c_p \gamma \Delta T}} = 0.0158 \text{ m}^2, \quad r = \rho\frac{2L}{s} = 9.34 \cdot 10^{-4} \text{ Ω}, \quad U_r = ir = 4670 \text{ V}. \tag{4}$$

where $r$ is electric resistance, Ohm; $U_r$ maximal electric intensity from Ohm resistance of rails, V.

4. *Maximal request voltage $U_m$*, average electric power $N$ and electric energy $A$ are:

$$U_m = E_m + U_r = 22.3 \cdot 10^3 \text{ V}, \quad N = 0.5iU_m = 5.5 \cdot 10^{10} \text{ W}, \quad A = Nt = 4.71 \cdot 10^9 \text{ J}, \tag{5}$$

The maximal electric power is two times more than 11 MkW. In our computation we neglect the loss of voltage in the generator.

5. *Rotor of the electric generator*. Radius of rotor the $n$ ($n$ = 20) studies electric generator $R$ for magnetic intensity $B$ = 1.2 T, and maximal the rotary speed $V_r$ = 600 m/s is

$$R = \frac{2U_m}{nBV_r} = 3.1 \text{ m} . \tag{6}$$

If one disk of rotor has mass of 200 kg, the total mass of rotor will be $m_r$ = 200×20 = 4000 kg.

6. *Fly-wheel*. Assume the fly-wheel made from artificial fiber having a safe tensile stress $\sigma$ = 100 kg/mm$^2$ = 10$^9$ N/m$^2$, specific density $\gamma$ = 2000 kg/m$^3$ . Then a safe rotary speed

$V_f = (\sigma/\gamma)^{0.5} = 710$ m/s and mass $M$ of fly-wheel is

$$M = 2A/V_f^2 + m_r = 20{,}000 + 4{,}000 = 24{,}000 \approx 25{,}000 \text{ kg} = 25 \text{ tons}. \tag{7}$$

We added 1 ton for friction in fly-wheel, reducer, generator, and loss in connection wire from electric generator to the rail. The maximal number of angular velocity is $\omega = V_f/R = 229$ radian/s = 36.5 revolution/s.

7. *Coefficient efficiency* of rails is $\eta_1 = E_m/U_m = 0.79$.

8. *Inductance $L_i$ and energy $W$* of a rail magnetic field are

$$L_i = \frac{\mu_o}{2}\left(0.5 + \ln\frac{d}{a}\right)L = 3.45 \cdot 10^{-4} \text{ H}, \quad W = \frac{iL_i}{2} = 4.33 \cdot 10^9 \text{ J}. \tag{8}$$

9. *Maximal repulsion force of rails* is

$$F_1 = \frac{\mu_o}{2\pi d}i^2 = 2.5 \cdot 10^7 \text{ N/m} = 2.5 \cdot 10^3 \text{ tons/m}. \tag{9}$$

This will merit your attention – this force is high and rails need in strong connection.

10. *Loss of launched apparatus speed in Earth's atmosphere* is about 100 m/s (see [1], pp.48-39, 135).

11. *Additional required fuel mass* to achieve delta-v at top of the trajectory (Fig.3) for circularization of the Earth orbit. (Typically a few hundred meters a second required velocity change.)

If rail angle is $\theta = 35^o$ degrees, the request orbit altitude is $H = 4000$ km and a solid rocket apparatus engine has impulse $w = 2000$ m/s, then request relative fuel mass is 0.2 (see [1], pp. 136-137).

If $\theta = 40 – 45^o$, $H = 400$ km, $w = 2280 – 3410$ m/s the request relative fuel is about 0.1. This means: from total mass of apparatus $m = 150$ kg the payload may be 100 -- 115 kg, the fuel 15 - 30 kg and the projectile body 15 - 35 kg.

12. *Cost of launch one kg of payload*. Assume the conventional turbo engine is used for moving the fly-wheel. Let us take the coefficient of efficiency the engine $\eta_1 = 0.3$ and the coefficient of efficiency of our launcher $\eta_2 = 0.6$. Then the requested amount of a fuel (gas, benzene) for one launch is

$$m_f = \frac{A}{\eta_1\eta_2 q} = 623 \text{ kg}, \tag{10}$$

where $A = 4.71 \times 10^9$ J is energy (see (5)), $q = 42 \times 10^6$ J/kg is heat capacity of fuel.

If cost of fuel is $c = \$0.5$/kg (end of 2008) and payload is 100 kg, the fuel cost per one kg is

$$C_f = c \cdot m_f / m_p = \$3.11/kg. \tag{11}$$

If frequency of launches is $t = 30$ min, the need power of engine is $N_e = A/\eta_2 t = 4.36 \times 10^3$ kW. That is power of a middle aviation turbo engine.

Let us assume the cost of magnetic launcher is 50 millions of dollars, lifetime of installation is 10 year and mountain is $2 millions of dollars per year. The launcher works the 350 days and launches 100 kg payload every 30 min (This means about 5000kg/day and 1750 tons/year). Then additional cost from installation is $C_i = \$2.86$/kg and total cost is

$$C = C_f + C_i \approx \$6/kg. \tag{12}$$

Compare this to the current cost of launching 1 kg of payload from $2500- $50000.

## Conclusion

The research shows the magnetic launcher can be built by the current technology. This significantly (by a thousand times) decreases the cost of space launches. Unfortunately, if we want to use the short rail way (412 m), any launcher request a big acceleration about $7.5 \cdot 10^3 g$ and may be used only for unmanned, hardened payload. If we want design the manned launcher the rail way must be 1100 km for acceleration $a = 3g$ (untrained passengers) and about 500 km ($a = 6g$) for trained cosmonauts.

Our design is not optimal. For example, the computation shows, if we increase our rail track only by 15 m, we do not need gas gun initial acceleration. That significantly decreases the cost of installation and simplifies its construction.

The reader can recalculate the installation for his own scenarios.

## Acknowledgement

The authors wish to acknowledge Joseph Friedlander (Israel) for correcting the author's English and for useful technical suggestions.